\documentclass[useAMS,usenatbib,a4paper]{mn2e}

\usepackage{aas_macros}

\usepackage{graphicx}
\usepackage{times}
\usepackage{amsmath}

\newcommand{\<}{\begin{eqnarray}}
\renewcommand{\>}{\end{eqnarray}} 
\renewcommand{\bar}{\overline}
\renewcommand{\tilde}{\widetilde}
\renewcommand{\hat}{\widehat}
\newcommand{\Msun}{\ensuremath{\rmn{M}_\odot}}
\renewcommand{\(}{\left(}
\renewcommand{\)}{\right)}
\renewcommand{\d}{\mathrm{d}} % for differentials

\newlength{\halfcolumn}
\newlength{\fullcolumn}
\newlength{\fullcolumnspace}
\providecommand{\numberLEQ}[1]{\refstepcounter{LEQCounter}\label{#1}\thetag{\theLEQCounter}}
\providecommand{\numberBEQ}[1]{\refstepcounter{BEQCounter}\label{#1}\thetag{\theBEQCounter}}

\title[On the function describing the IMF.]{
On the function describing the stellar initial mass function.
}
\author[Th. Maschberger]
{Th. Maschberger$^{1}$\thanks{e-mail: thomas.maschberger@obs.ujf-grenoble.fr}
\\
\small \it 
$^1$ Institut de Plan{\'e}tologie et d{'}Astrophysique de Grenoble, BP 53, F-38041 Grenoble C{\'e}dex 9, France\\
}
\date{}

\pagerange{\pageref{firstpage}--\pageref{lastpage}} \pubyear{2011}

\begin{document}
\label{firstpage}
\newcounter{LEQCounter}
\newcounter{BEQCounter}

\maketitle

\begin{abstract}
We propose a functional form for the IMF, the $L_3$ IMF, which is a natural heavy-tailed approximation to the log-normal distribution.
It is composed of a low-mass power law and a high mass power-law which are smoothly joined together. 
Three parameters are needed to achieve this.
The standard IMFs of \citet{Kroupa-2001a,Kroupa-2002a} and \citet{Chabrier-2003b} (single stars or systems) are essentially indistinguishable from this form.
Compared to other 3-parameter functions of the IMF, the $L_3$ IMF has the advantage that the cumulative distribution function and many other characteristic quantities have a closed form, the mass generating function, for example, can be written down explicitly.
\end{abstract}


\section{Introduction}
The initial mass function of stars (IMF), the spectrum of stellar masses at their birth, is of  fundamental importance in many fields of Astronomy.
Since the seminal work of \citet{Salpeter-1955a}, who investigated the power-law part of the massive stars, a huge observational and theoretical effort has been made to constrain this distribution.
Towards the lesser masses the IMF deviates from a power law and follows more a lognormal shape \citep{MillerScalo-1979a}.
At present, the whole shape of the IMF is usually described by power-law segments \citep{Kroupa-2001a,Kroupa-2002a} or by a lognormal segment plus a power law segment \citep{Chabrier-2003b,Chabrier-2003a,Chabrier-2005a}.
The aim of this paper is to provide an alternative, practical functional form for the IMF together with all its characteristic quantities (see Table \ref{formulae_L3} for the formulae and Figs.  \ref{figure_imf_L3} and \ref{figure_imf_loglog_L3})
\footnote{R code for the functions given in this paper is available as online material}
.
More observational and theoretical aspects of the IMF can be found in recent reviews \citep[e.g.][]{Scalo-1986a,Chabrier-2003b,ZinneckerYorke-2007a,Elmegreen-2009b,BastianCoveyMeyer-2010b,KroupaWeidnerPflamm-Altenburg-2011b}.

The IMF is usually believed to be a smooth function over the whole mass range, from brown dwarfs to O stars.
However, \citet{ThiesKroupa-2007a} and \citet{ThiesKroupa-2008a} argued that a sudden change in binarity properties around the hydrogen burning limit introduces a discontinuity in the single star IMF as well.
This discontinuity in the single-star IMF can still lead to a system IMF without discontinuities over the whole mass range 
\citep{ThiesKroupa-2007a,KroupaWeidnerPflamm-Altenburg-2011b}.
In view of the simplicity aspect of our proposed IMF form we neglect any discontinuity.

The proposed functional form, the $L_3$ IMF, fulfils several demands on the form of the IMF:
It describes the whole (system) mass range with a single function.
This has been achieved by several other functional forms as well \citep{Larson-1998a,Chabrier-2001a,Parescede-Marchi-2000a,ParravanoMcKeeHollenbach-2011a,CartwrightWhitworth-2012a}.
However, compared to these forms, the $L_3$ IMF has the advantage that its cumulative distribution function is invertible, so that sampling from the $L_3$ IMF is very easy.
No special functions (e.g. the error function) are involved to normalise the $L_3$ IMF as a probability.
Beyond that the analytical form allows also for simple, closed forms of characteristic quantities, such as the peak or the ``breaks'', the masses from which on the power laws reigns.
Furthermore, with three parameters, two controlling the power-law behaviour at low and high masses and one location parameter, the number of parameters is as small as possible.

The motivation for the $L_3$ IMF is of purely pragmatic nature, it is a functional form that describes the data in a very practical way.
It would be pleasing if the $L_3$ IMF could be more ``theoretically'' motivated.
One could try to find a connection to some generalised log-logistic growth processes, in analogy to logistic growth, as the $L_3$ is related to the log-logistic distribution.
However, it remains questionable whether such a (non-stochastic) growth theory would be capturing the star formation process in its entirety \citep[cf. the discussion about logistic growth in ][ p. 52]{Feller-1968b}.
Where would be the place of, for example, feedback or stellar dynamics in shaping the IMF if growth alone 
gives all parameters of the IMF?
Thus it seems futile to follow such thoughts and we do not attempt to find any reasons for our proposed functional form, other than its utmost simplicity and practicality.

The organisation of this paper is the following:
After some general definitions we discuss in Section \ref{sec_properties_IMF} established functional forms and required parameters of the IMF.
The $L_3$ and $B_4$ IMFs are motivated and defined in Section \ref{sec_heavytails} as heavy-tailed approximations and extensions to the log-normal distribution.
This is followed by a detailed description of the $L_3$ IMF and its characteristic quantities in Section \ref{sec_L3}, the $B_4$ IMF is discussed in Appendix \ref{B4_IMF}.
Section \ref{sec_std_IMF} gives the ``canonical'' parameters for the $L_3$ IMF, matching it to the \citet{Kroupa-2001a,Kroupa-2002a} and \citet{Chabrier-2003b,Chabrier-2003a} IMFs.
Sec. \ref{sec_summary} contains the  conclusions of this article.

\section{Properties of the IMF}\label{sec_properties_IMF}

\subsection{Definitions}\label{sec_def}
We normalise the IMF as a {\it probability density function} (pdf), the IMF tells us about the relative frequencies of stars of various masses in linear mass space.
This allows us to use common statistical techniques, e.g. to estimate the parameters.
For functions normalised as pdf we use the symbol $p (m)$, for their integrals, the cumulative distribution function, the symbol $P(m)$.
The cumulative distribution function is related to the observed number frequency, $N(m)$, by
$P(m) = \frac{1}{n_\text{tot}} N(m)$, where $n_\text{tot}$ is the total number of observed stars.
The standard normalisation condition for a probability is
\< 1 &=& \int_{m_l}^{m_u} p(m) \d m, \>
where $m_l$ and $m_u$ are the lower and upper mass limit, respectively.

Historically there exist two alternative descriptions of the IMF, in linear or in logarithmic space,  the small-$\alpha$ and the big-$\Gamma$ notation.
The use of the IMF as probability of $m$ leads naturally to the linear (small-$\alpha$) description, the IMF is fulfils
\< p(m) &=& \frac{\d P(m)}{\d m}  \left[ = \frac{1}{n_\text{tot}} \frac{\d N(m)}{\d m} \right] .\>
A power law IMF has then the exponent $-\alpha$, $p(m) \propto m^{-\alpha}$.
In the logarithmic description the IMF is normalised as probability of $\log m$, not $m$,
\< p_\text{log} ( \log m) &=& \frac{\d P_\text{log} (\log m)}{\d \log m}  \left[ = \frac{1}{n_\text{tot}} \frac{\d N_\text{log} (\log m) }{\d \log m } \right]. \>
$p_\text{log} (\log m) $ is connected to the linear pdf via
\< p (m) &=& \frac{\d P_\text{log} (\log m) }{\d \log m}  \frac{\d \log m}{\d m} 
= \frac{1}{m} p_\text{log} (m)    .
\>
Thus, a power law pdf in $m$, $p(m) \propto m^{-\alpha}$, transforms into
$p_\text{log} (m) \propto m^{-(\alpha -1)}$ or $p_\text{log} (m) \propto m^{-\Gamma}$, where
$\Gamma = \alpha -1$.

We define the exponent (sometimes referred to as ``slope'', but that should be reserved for the logarithmic description), as a function of mass via
\< S(m) &=& - \frac{\d \log p(m)}{\d \log m} = - m \frac{\d \log p(m)}{\d m}. \label{def_slope} \>
A power-law IMF can then be written as
\< p(m) &\propto& m^{- S(m)}. \>
We follow the convention that the negative sign is not included in the exponent.
Thus, in our notation the \citet{Salpeter-1955a} exponent is positive, $\alpha =${\bf +}2.35.

\subsection{The standard IMFs and other  functional forms}
The \cite{Kroupa-2001a,Kroupa-2002a} single-star IMF consists only of power-law segments,
\< p_\text{Kroupa} (m ) & = &  \begin{cases}
\begin{array}{lr@{}l@{ }c@{ }c@{\ }r@{}l}
A k_0 m^{- 0.3} &  0&.01\ \Msun &< m &<& 0&.08\  \Msun \\
A k_1 m^{- 1.3} &  0&.08\ \Msun  &< m& < & 0&.5\  \Msun \\
A k_2 m^{- 2.3} &  0&.5 \ \Msun &< m & < &1&\ \Msun \\
A k_3 m^{- 2.3} &  1& \ \Msun &< m & \multicolumn{3}{c}{ ( < 150 \ \Msun)}
\end{array},
\end{cases}
\label{IMF_kroupa}
\>
with
$k_0 = 1$, $k_1 = k_0 m_1^{-0.3 + 1.3}$, $k_2 = k_1 m_2^{-1.3 + 2.3}$ and $k_3 = k_2 m_3^{-2.3 + 2.3} (=k_2)$ where $m_1=0.08\ \Msun$, $m_2=0.5\ \Msun$ and $m_3 = 1\ \Msun$
\citep[a practical algorithm for the calculation of the $k_i$ is given by ][]{Pflamm-AltenburgKroupa-2006a}.
$A$ is some global normalisation constant.
This form is highly adaptable, which comes at the price of a large number of parameters.
On the practical side, the \cite{Kroupa-2001a,Kroupa-2002a} IMF has the advantage that many derived quantities can be calculated without involving special functions (cumulative distribution function, quantile function, mean mass etc.), but with several ``if'' statements to specify the mass ranges.

\citet{Chabrier-2003b,Chabrier-2003a} combined for the single-star IMF a log-normal distribution at the low-mass end with a high-mass power law,
\< p_\text{Chabrier} (m) =  \begin{cases}
\begin{array}{l@{}r@{\ }c@{\ }l}
\!\!\! A k_1 \frac{1}{m} e^{-  \frac{1}{2} \(  \frac{ \log_{10} m - \log_{10} 0.079}{0.69}\)^2 }
 &  m  & <  1 \ \Msun \\
\!\!\! A  k_2 m^{-2.3} & m & >    1 \ \Msun 
\end{array}
\end{cases},
\label{IMF_chabrier}
\>
with $k_1 = 0.158$ and $k_2 = 0.0443$ and the global normalisation constant $A$.
The lognormal and the power-law part connect up more or less smoothly, without the ``kinks'' of several power-law segments (although there is still the small kink at 1\ \Msun).
Calculating the cumulative distribution function involves the error function, but random variates can be created without any specialised algorithms from standard Gaussian distributed random numbers.

A piece-wise functional form of the IMF is somewhat unsatisfying, and several alternatives covering the whole mass range have been proposed in the literature.
There are, for example, the functional forms of 
\citet{Larson-1998a}
\< p_\text{Larson a} (m) &\propto&  \frac{1}{m} \(1+\frac{m}{\mu} \)^{-(\alpha+1)} \label{larson_a}, \>
and 
\< p_\text{Larson b} (m) &\propto&  m^{-\alpha} e^{- \( \frac{m}{\mu} \)^{-1} }  \label{larson_b}, \>
form 3 of \citet{Chabrier-2001a},
\< p_\text{Chabrier 3} (m) &\propto& m^{-\alpha} e^{- \( \frac{m}{\mu} \)^{-\beta} }, \label{chabrier_3} \>
or the 
tapered power law form of  \citet{Parescede-Marchi-2000a}, \citet{De-MarchiParescePortegies-Zwart-2010a}, \citet{HollenbachParravanoMcKee-2005a} and \citet{ParravanoMcKeeHollenbach-2011a},
\< p_\text{Taperered PL} (m) & \propto &  m^{-\alpha} \( 1 - e^{- \(\frac{m}{\mu}\)^{- \beta} }\) \label{stpl}. \>
The IMF forms of eqq. \ref{larson_a}, \ref{larson_b}, \ref{chabrier_3} and \ref{stpl} are very similar to our proposed form of the IMF, but their integrals contain the incomplete gamma function or the hypergeometric function.
A cumulative distribution function without closed form is hard to invert, so that special algorithms are necessary for random variates from these distributions.

Recently, \cite{CartwrightWhitworth-2012a} proposed a completely different class of distribution functions for the IMF description, stable distributions.
Stable distribution (e.g., the Gaussian distribution) arise naturally in the context of stochastic processes, of which the star formation process is one example.
Related to stable distributions, and also the outcome of stochastic processes is the class of infinitely divisible distributions, such as the lognormal distribution (e.g. \citealp{Zinnecker-1984a}, \citealp{ElmegreenMathieu-1983a};  \citealp{Thorin-1977a} for infinite divisibility).
The choice of stable distributions is motivated by their relation to stochastic processes, however, they are also used only as a fitting function, as the exact stochastic process describing star formation has not yet been formalised.
Also, typically they do not have a closed form for the distribution function itself, which is an important practical aspect.

\subsection{How many parameters for the IMF?}
The IMF seems to have a lognormal body with a power law tail on both the high-mass and the low-mass side.
In order to describe this behaviour, four parameters appear to be required: a location parameter (which is not necessarily the ``peak'' or the mean), a scale or width parameter (which is not necessarily the variance), the low-mass and high-mass power-law exponents.
There are no stars of zero or infinite mass, so that additionally an upper and a lower mass limit has to be introduced, so the total number of parameters is 4+2.
This is two parameters less than in the schematic IMF of \citet{BastianCoveyMeyer-2010b}, where additionally two ``mass breaks'' are introduced, i.e. 6+2 parameters.
However, if one requires that the lognormal part merges smoothly into the power law tails, then the scale parameter sets the width of the IMF and consequently the mass breaks.
The mass ``breaks'' are then not parameters any more, but derived quantities.
4+2  seem therefore to be the necessary number of parameters to describe the IMF.
The $B_4$ IMF discussed later is a smooth function over all masses and has the mentioned 4+2 parameters.

The number of parameters of the IMF can be reduced by one, because it is not necessary to explicitly include a scale parameter to fit the ``canonical'' IMF.
Only a location parameter and the two exponents suffice to achieve this.
Several 3+2 IMFs have been suggested in the literature (eq. \ref{chabrier_3}, IMF 3 of \citealp{Chabrier-2001a}; eq. \ref{stpl} \citealp{Parescede-Marchi-2000a,De-MarchiParescePortegies-Zwart-2010a,HollenbachParravanoMcKee-2005a,ParravanoMcKeeHollenbach-2011a}).
Our proposed $L_3$ IMF also has only 3+2 parameters.

2+2 parameter functional forms (eq. \ref{larson_a} and \ref{larson_b}) have been given by \citet{Larson-1998a}, with a location parameter and only a high-mass exponent.
With only 2+2 parameters it is difficult to fit the low-mass end of the IMF.

For comparison, the \citet{Kroupa-2001a,Kroupa-2002a}
has 5+2 parameters (three exponents, two thresholds, two limits) and the \citet{Chabrier-2003b,Chabrier-2003a} IMF
has 4+2 parameters (mean, variance, one exponent, one threshold, two limits).

\section{Heavy-tailed approximations to the lognormal distribution}\label{sec_heavytails}
\begin{figure}
\begin{center}
\includegraphics[width=6cm]{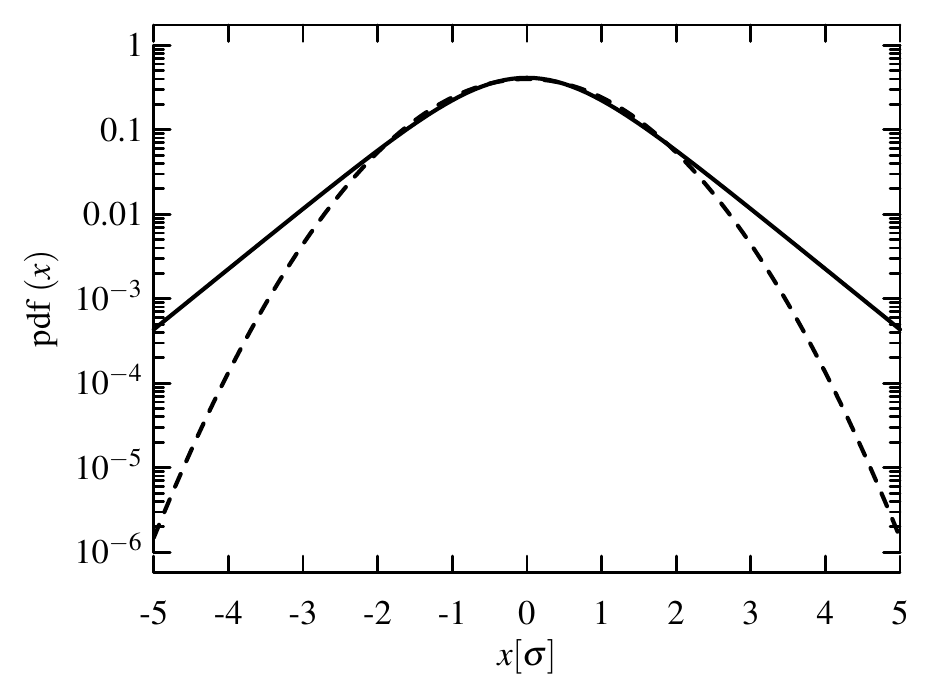}
\end{center}
\caption{\label{normal_logistic}
Comparison of  the Normal (dashed line) and Logistic (solid line) probability density, with a logarithmic $y$ axis.
The Logistic distribution has heavier tails.
}
\end{figure}

Starting point for the search of a functional form for the IMF is the relation between the Normal distribution and the Logistic distribution \citep[see e.g.][]{JohnsonKotzBalakrishnan-1994a,JohnsonKotzBalakrishnan-1995a}.
The Normal distribution,
\< p_\mathcal{N} (x) &=& \frac{1}{\sqrt{2 \pi}\sigma} e^{-\frac{1}{2} \frac{(x-\mu)^2}{\sigma^2} }, \>
can be approximated in the central region for $\sigma = 1 $ by the Logistic distribution,
\< p_L (x) &=& \frac{1}{\sigma'} \frac{e^{- \frac{x-\mu}{\sigma'}}}{\( 1 + e^{- \frac{x-\mu}{\sigma'}}\)^2 }, \>
where $\sigma' = e^{-\frac{1}{2}}$.
The ratio of the two probability densities is close to unity between $-2 \sigma$ and $+ 2 \sigma$, but drops off strongly outside.
This behaviour is evident in a logarithmic plot of both densities (Fig. \ref{normal_logistic}), the tails of the logistic distribution are much heavier than the normal distribution, with fixed exponents.

\begin{figure}
\begin{center}
\includegraphics[width=6.5cm]{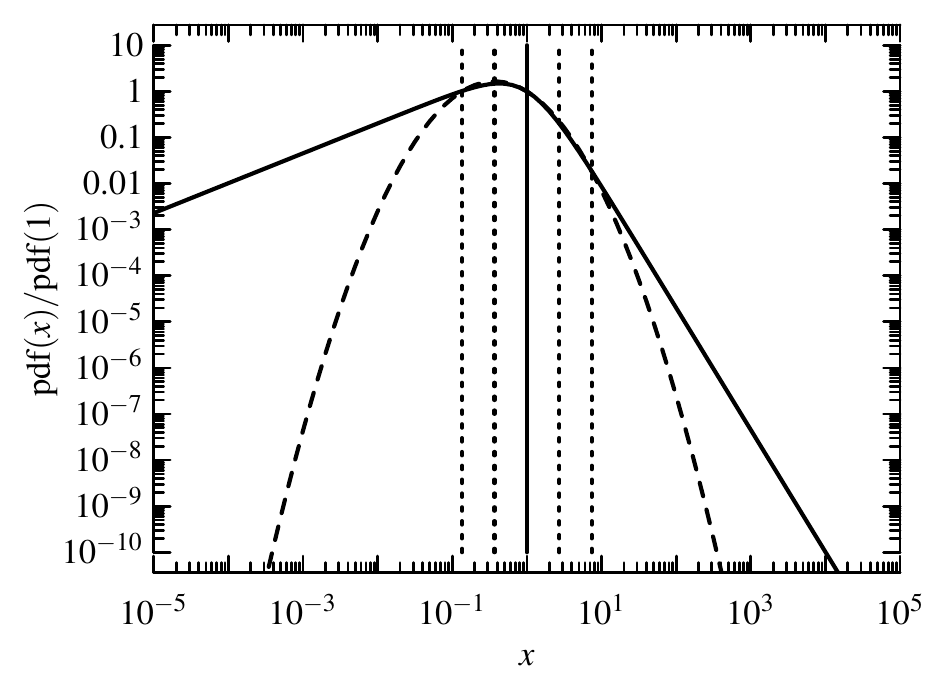}
\end{center}
\caption{\label{lognormal_loglogistic}
Comparison of the log-Normal (dashed line) and  log-Logistic (solid line) distribution, scaled to $1$ at $x=1$.
The dotted lines are at $e^{-2}$, $e^{-1}$, $e^{1}$ and $e^{2}$.
The tails of the log-Logistic distribution are asymmetric.
}
\end{figure}

\begin{table*}
\caption{\label{formulae_L3}
Collection of formulae for the $L_3$ form of the IMF.
The values given for the parameters and characteristic quantities are to match the ``canonical'' single-star IMF \citep{Kroupa-2001a,Kroupa-2002a,Chabrier-2003b}, values in parentheses for the ``canonical'' system (binary star) IMF \citep{Chabrier-2003b}.
$B (t;p,q)$ is the incomplete Beta function.
For the limits we adopt the fiducial values $m_l=0.01\ \Msun$ and $m_u = 150\ \Msun$, which are only needed for the normalisation.
}

\setcounter{LEQCounter}{0}
\begin{tabular}{clll}
\numberLEQ{L3_aux}&
Auxilliary function: &
$ \displaystyle G (m) = 
%  B_0^\infty \( \(\frac{m}{\mu}\)^{\frac{1}{\sigma}}; \sigma (\beta + i) , \sigma (\alpha - i) \) =
 \( 1 + \( \frac{m}{\mu} \)^{1-\alpha} \)^{1-\beta}$
&
\\[3em]
&\bf Quantity & \bf Formula & \bf Definition \\
&\multicolumn{3}{l}{\bf Functional form}\\
\numberLEQ{L3_CDF}&
Cumulative distribution function (CDF) &
$ \displaystyle P_{L3} (m) = \frac{G(m) - G(m_l)}{G(m_u) - G(m_l)}$  & 
$ \displaystyle P (m_l) = 0$ and $P(m_u) = 1$
\\[1em]
\numberLEQ{L3_pdf}&
Probability density function (pdf) &
$ \displaystyle p_{L3} (m)= A \( \frac{m}{\mu} \)^{-\alpha} \( 1 + \( \frac{m}{\mu} \)^{1-\alpha} \)^{-\beta} $ &
$ \displaystyle p(m) = \frac{\d }{\d m} P(m) $
\\[2em]
&
&
$ \displaystyle A = \frac{(1-\alpha)(1-\beta)}{\mu}  \frac{1}{G(m_u) - G(m_l)}$ & 
\\[2em] 
\numberLEQ{L3_QF}&
Quantile function &
$ \displaystyle m(u) =  \mu \( \Bigg[ u \Big( G (m_u) - G(m_l) \Big) + G(m_l) \Bigg]^{\frac{1}{1-\beta}} - 1  \)^{\frac{1}{1-\alpha}} $ &
$ \displaystyle m(u) = P^{-1} (u) $, $u \in [0,1]$
\\[1em]
&\multicolumn{3}{l}{\bf Parameters }\\
&
High-mass exponent &
$\alpha = 2.3\ (2.3)$ &
$\alpha \neq 1 $ (typically $\alpha > 0$) 
\\
&
Low-mass exponent &
$\beta = 1.4 \ (2.0)$&
$\beta \neq 1$ (typically $\beta > 0 $) 
\\
&
Scale parameter &
$\mu = 0.2\ (0.2) \ \Msun $ & 
$\mu > 0$
\\
&
Lower mass limit &
$m_l = 0.01\ \Msun $ &
$m_l > 0$
\\
&
Upper mass limit & 
$m_u=  150\ \Msun $ &
$m_u > 0$
\\ 
&\multicolumn{3}{l}{\bf Shape characterising quantities }\\
&
Effective high-mass exponent &
$ \alpha = 2.3 \ (2.3)$ &
$ \displaystyle \lim_{m \to \infty} p(m) \propto m^{-\alpha}$
\\
\numberLEQ{L3_gamma}&
Effective low-mass exponent &
$ \gamma =  \alpha + \beta ( 1 - \alpha) = 0.48\ (-0.3)$ &
$ \displaystyle \lim_{m \to 0} p(m) \propto m^{-\gamma}$
\\
\numberLEQ{L3_mgamma}&
Lower power-law mass limit &
$ \displaystyle m_\gamma = \mu e^{\frac{2}{1-\alpha}} =0.043\ (0.043)\ \Msun $ & 
$ \displaystyle p(m \in [m_l \dots m_\gamma]) \approx m^{-\gamma}$
\\
\numberLEQ{L3_malpha}&
Upper power-law mass limit &
$ \displaystyle m_\alpha = \mu e^{\frac{2}{\alpha-1}} =  0.93\ (0.93)\ \Msun $ & 
$ \displaystyle p(m \in [m_\alpha \dots m_u]) \approx  m^{-\alpha}$ % \stackrel{\text{D}}{\approx}
\\
\numberLEQ{L3_slope}&
Exponent (N.B. $S(\infty) = \alpha$ {\tiny $(=  2.3)$})  & 
$ \displaystyle S(m) =  \alpha + \beta (1-\alpha)  \( 1+ \(\frac{m}{\mu}\)^{1-\alpha} \)^{-1}  \(\frac{m}{\mu}\)^{1-\alpha} $ &
$ \displaystyle S(m) = - \frac{\d \log p(m)}{\d \log m}   $
\\[1em] 
&\multicolumn{3}{l}{\bf Scale characterising quantities }\\
&Mean mass (Expectation value) &
\begin{minipage}[t]{7cm}
$ \displaystyle E(m) \( = \bar{m} \)= $
expressed by Beta function, see eq. \ref{mean_L3}\\
\hspace*{1.35cm} $=0.36\  (0.62)\ \Msun$ \end{minipage} &
$ \displaystyle E(m) = \int_{m_l}^{m_u} m p (m ) \d m$
\\[1em]
\numberLEQ{L3_median}&
Median mass &
\begin{minipage}[t]{7cm}
$ \displaystyle \tilde{m} =    \mu \( \Bigg[ \frac{1}{2} \Big( G (m_u) - G(m_l) \Big) + G(m_l) \Bigg]^{\frac{1}{1-\beta}} - 1  \)^{\frac{1}{1-\alpha}}$ \\
\hspace*{.22cm} $   = 0.10\ (0.21)\ \Msun$ \end{minipage}&
$ \displaystyle P(\tilde{m}) = \frac{1}{2} $
\\[2em]
\numberLEQ{L3_mode}&
Mode (most probable mass)
&
$\hat{m} =
\begin{cases}
\mu \(  \frac{\beta (\alpha -1 )}{\alpha} - 1 \)^\frac{1}{(\alpha-1)}  & \gamma < 0 \\
 m_l & \gamma > 0 
 \end{cases}  = 0.01\ (0.04)\ \Msun $
&
$ \displaystyle \hat{m} = \underset{m}{\mathrm{arg\ max}} p(m) $
\\[1em]
\numberLEQ{L3_peak}&
``Peak'' (maximum in log-log) 
&
$ \displaystyle m_P = \mu \( \beta- 1 \)^{\frac{1}{\alpha-1}} = 0.10\ (0.20)\ \Msun$ &
$ \displaystyle \frac{\d \log \( m p(m) \)}{\d \log m}  = 0 $
\\
\end{tabular}

\end{table*}

In order to translate the relation of Normal and Logistic distribution to the lognormal distribution, 
\< p_{\log\mathcal{N}} (x) &\propto& \frac{1}{x} e^{ - \frac{1}{2} \(  \frac{\ln x - \ln \mu}{\sigma} \)^2}, \>
we rewrite the lognormal density function as 
\< p_{\log\mathcal{N}}  (x) &\propto& \frac{1}{x} e^{- \frac{1}{2} \( \ln \left[ \(\frac{x}{\mu}\)^{\frac{1}{\sigma}} \right] \)^2 }. \>
Inserting $  \ln \left[ \(\frac{x}{\mu}\)^{\frac{1}{\sigma'}} \right] $ for $\frac{x-\mu}{\sigma'}$ into the Logistic cumulative distribution function,
\< P_{\log  L} (x) &=& \frac{1}{1+e^{- \frac{x-\mu}{\sigma'} }}, \>
and taking the derivative gives the log-Logistic density,
\< p_{\log L} (x)  &\propto&  \frac{ \(\frac{x}{\mu}\)^{-\frac{1}{\sigma'}-1} }{ \( 1+ \(\frac{x}{\mu}\)^{-\frac{1}{\sigma'}} \)^2 }. \>
Figure \ref{lognormal_loglogistic} shows $p_{\log\mathcal{N}}$ and $p_{\log L}$, again with $\sigma=1$ and $\sigma' = e^{-\frac{1}{2}}$.
The log-logistic distribution follows the lognormal distribution over about two orders of magnitude and deviates with asymmetric tails.

The log-Logistic distribution of Fig. \ref{lognormal_loglogistic} already looks very much like the IMF.
Only the high-mass and low-mass exponents are still fixed.
In fact, this is not quite correct, because the meaning of $\sigma'$ has been changed from the width of the distribution (i.e. a scale parameter) to determining the low-mass exponent (i.e. a shape parameter).
Arbitrary exponents for the low-mass and the high-mass tail can be introduced by writing
\< p_{L3} (m) &\propto& \frac{ \( \frac{m}{\mu}\)^{-\alpha} }{ \( 1 + \( \frac{m}{\mu}\)^{1-\alpha} \)^{\beta}}. \label{pdf_L3} \>
Unfortunately, $\beta$ is {\it  not} the exponent at low masses, which is the price paid for eq. \ref{pdf_L3} having a very simple cumulative distribution function.
Probability densities similar to eq. \ref{pdf_L3} (two exponents and $\mu$)  are known under several other names, particularly in economics.
We will refer to it as generalised log-Logistic distribution, or in short ``$L_3$ IMF'', because it has three (shape) parameters.

A parameter that changes the width of the IMF can be introduced by writing
\< p_{B4} (m) &=& \frac{ \(\frac{m}{\mu} \)^{\beta}}{ \( 1 + \( \frac{m}{\mu} \)^\frac{1}{\sigma} \)^{\sigma (\alpha + \beta)}}. \label{pdf_B4}\>
$\sigma$ is now the scale parameter, and $\alpha$ and $\beta$ the exponents of the power-law tails.
The integral of eq. \ref{pdf_B4} does not have a closed form, but can be transformed to the incomplete Beta function.
Therefore, probability densities of the type of eq. \ref{pdf_B4} are known as (generalised) Beta distributions.
Because of the four parameters we will refer to it as $B_4$ IMF.

The following Sections will show, that the ``canonical'' IMF \citep{Kroupa-2001a,Kroupa-2002a,Chabrier-2003b} can be very satisfyingly described by the $L_3$ IMF.
The introduction of $\sigma$ as an additional scale parameter seems not to be necessary.
Therefore we consider in the following only the $L_3$ IMF and give the corresponding equations and parameter values for the $B_4$ IMF in appendix \ref{B4_IMF}.

\section{The $L_3$ IMF}\label{sec_L3}

\subsection{Functional form}
\begin{figure}
\includegraphics[width=8cm]{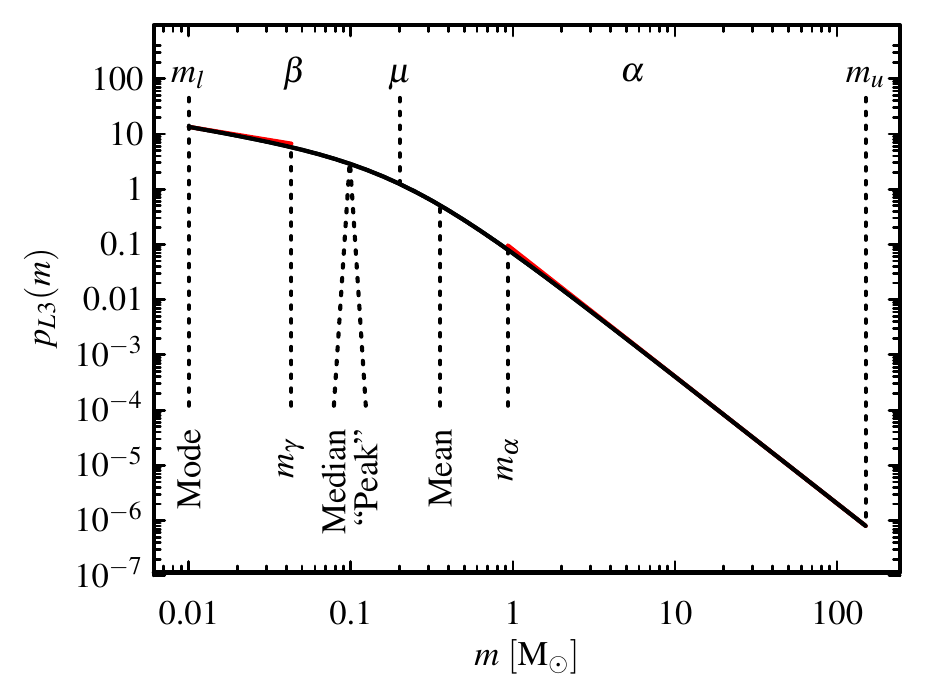}
\caption{ \label{figure_imf_L3}
Probability density function for the $L_3$ functional form of the IMF (Table \ref{formulae_L3} eq. \ref{L3_pdf}) with the ``canonical'' parameters, $\alpha=2.3$, $\beta=1.4$, $\mu=0.2$ and the limits $m_l=0.01\ \Msun$ and $m_u = 150\ \Msun$.
It follows approximately  $m^{-\alpha}$ for $m > m_\alpha (\approx 0.9\ \Msun)$ and $m^{-\gamma} = m^{-\(\alpha + \beta(1-\alpha)\)}$ for $m < m_\gamma (\approx 0.04\ \Msun)$ with $\gamma = 0.48$.
Also shown are the locations of mean, median and mode, which are all different because of the skewed distribution.
The infamous ``peak'' (maximum in log-log) {\it is not} the location at which the two power laws cross over. 
This happens at the scale parameter $\mu$.
}
\end{figure}

The probability density of the $L_3$ IMF is given in eq. \ref{pdf_L3}, or, with the normalisation constant, in Table \ref{formulae_L3} eq. \ref{L3_pdf}.
Table \ref{formulae_L3}  collects all formulae for the $L_3$ IMF.
Figure \ref{figure_imf_L3} shows the $L_3$ IMF with its characteristic quantities for the ``canonical parameters'' of the single-star IMF.
The particular advantage of the $L_3$ IMF is that the integral of the probability density is very simple,
\<  \int  \frac{ \( \frac{m}{\mu}\)^{-\alpha} }{ \( 1 + \( \frac{m}{\mu}\)^{1-\alpha} \)^{\beta}} \d m
\propto  \( 1 + \( \frac{m}{\mu}\)^{1-\alpha} \)^{1-\beta}
\!\!\!\!\!\!\! =: G(m)
. \>
The full cumulative distribution function, including the upper and lower limits ($m_l$ and $m_u$), is then
\< P(m) &=& \frac{G(m) - G(m_l)}{G(m_u) - G(m_l)} \label{CDF_L3} \>
(also eq. \ref{L3_CDF}, Table \ref{formulae_L3}).
Eq. \ref{CDF_L3} can be readily inverted to give the quantile function (Eq. \ref{L3_QF}, Table \ref{formulae_L3}).
Generating a random mass from the $L_3$ IMF (i.e. inserting a uniform random number $u$ in the quantile function) can then essentially be done in a single line of code.
% (perhaps not {\sc fortran77}).

\begin{figure}
\begin{center}
\includegraphics[width=7cm]{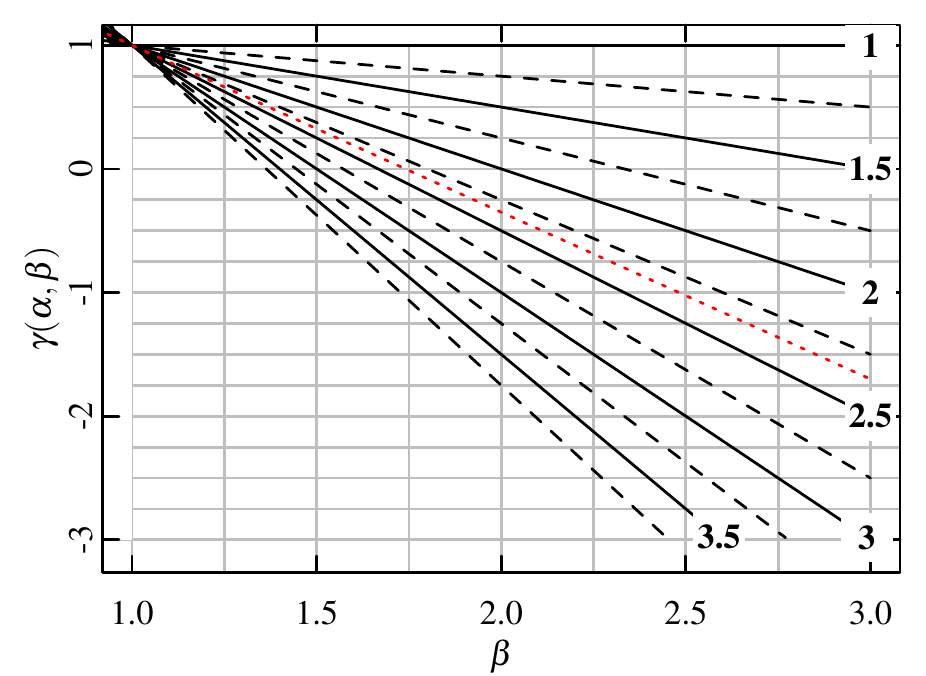}
\end{center}
\caption{\label{alphabetagamma}
$\alpha\beta\gamma$ plot,  showing the value of the low-mass exponent ($p(m) \propto m^{-\gamma}$, Table \ref{formulae_L3} eq. \ref{L3_gamma}) as a function of $\alpha$ and $\beta$. 
The lines solid for integer $\alpha$ and $\alpha+1/2$ and dashed for $\alpha+1/4$ and $\alpha+3/4$.
The red dotted line is for $\alpha=2.35$.
}
\end{figure}

The two shape parameters have different meanings for the $L_3$ IMF.
For large masses $\lim_{m \to \infty} p(m) \propto m^{-\alpha} $, i.e. $\alpha$ is the high-mass exponent.
In order that the $L_3$ IMF is defined $\alpha \neq 1$ is required, typically will be $\alpha > 1$.
For small masses the limiting case is $\lim_{m \to 0} p(m) \propto m^{-\gamma}$ with $\gamma = \alpha + \beta ( 1 - \alpha)$.
Therefore the parameter $\beta$ {\it is not} the low-mass exponent.
This inconvenience of $\beta$ and $\gamma$ is the trade-off for the very simple cumulative distribution, 
Again, in order for the $L_3$ IMF to be defined $\beta \neq 1$ is required, typically will be $\beta > 1$.
For $\alpha >1 $ and $\beta > 1$ the largest value that $\gamma$ can take is $+1$, i.e. $p(m) \propto m^{-1}$.
$\gamma$ will be negative for $\beta > \frac{\alpha}{\alpha-1}$.
A graphical representation of the relation between the exponents is given in the ``$\alpha\beta\gamma$ plot'', Fig. \ref{alphabetagamma}, where the value of $\gamma$ for given $\alpha$ and $\beta$ can easily be read off.

\begin{figure}
\includegraphics[width=8cm]{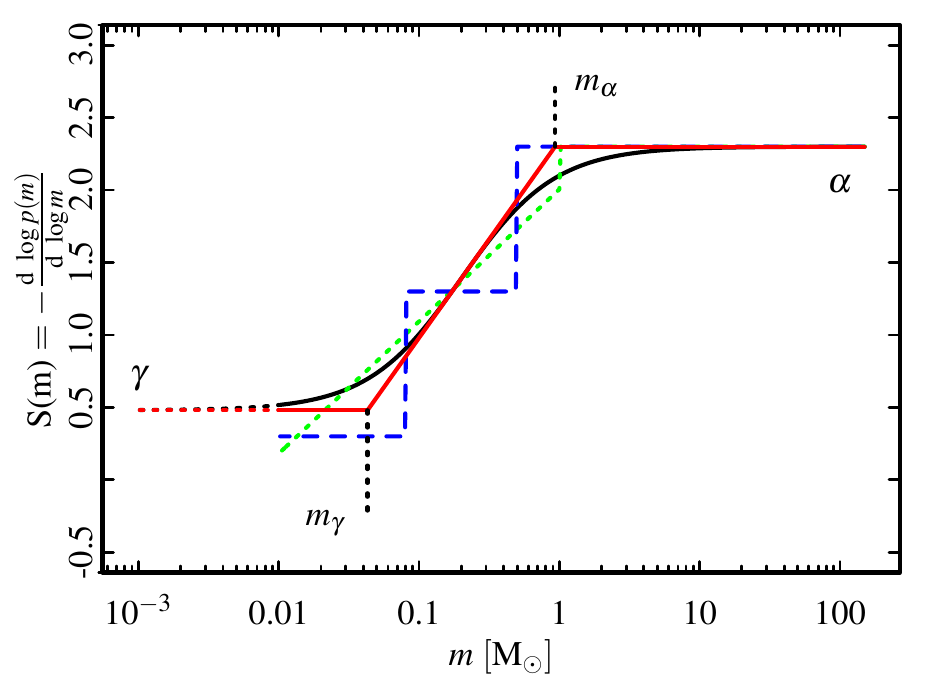}
\caption{ \label{figure_slope_L3}
(``Alpha plot'') Exponent of the $L_3$ IMF (black solid curve) and its approximation in log space (red solid lines).
The points $m_\alpha$ and $m_\gamma$ are defined as the intersection of the straight line approximation of the exponent at $\mu$ with the limiting exponents $\gamma$ and $\alpha$.
For comparison the \citet{Kroupa-2001a,Kroupa-2002a} IMF (blue dashed line) and  the \citet{Chabrier-2003b} IMF (green dotted line) are given as well.
}
\end{figure}

\subsection{Breakpoints}
Related to the low- and high-mass exponents is the question of the ``breakpoints'' in the IMF.
As for the $L_3$ (and the $B_4$) IMF there is a smooth transition between the exponents, proper breakpoints do not exist.
Nevertheless, it is useful to know from where the $L_3$ IMF can be approximated by a power law.
Our approach to find the breakpoints is via the exponent as function of mass (defined in eq. \ref{def_slope}, given for the $L_3$ IMF in eq. \ref{L3_slope}, Table \ref{formulae_L3})
For the $L_3$ IMF the curve of the exponent vs. $\log m$ is  ``S''-shaped, see the black solid line in Fig. \ref{figure_slope_L3}.
This ``S'' shape can be approximated by three straight lines (red in Fig. \ref{figure_slope_L3}), of which two are horizontal at $\gamma$ and $\alpha$.
The intermediate, increasing part follows
\< g(m) &=&  \left. \frac{\d S(m) }{\d \log m}\right|_\mu \log \( \frac{m}{\mu} \) + S(\mu), \>
a straight line in $\log m$.
We define now the breakpoints, $m_\gamma$ and $m_\alpha$, as the points where $g(m_\gamma)=\gamma$ and $g(m_\alpha)=\alpha$.
Formulae are given in Table \ref{formulae_L3}, eqq. \ref{L3_mgamma} and \ref{L3_malpha}.
The agreement of the $L_3$ IMF and the power-law segments below $m_\gamma$ and above $m_\alpha$ is good, as can be seen in Fig. \ref{figure_imf_L3}, where the power-law segments are shown as red lines, which are in fact barely visible.

\subsection{Characteristic masses}
Characteristic mass scales of the $L_3$ IMF are also shown in Fig. \ref{figure_imf_L3} and given in Table \ref{formulae_L3}.
Because the IMF is skewed, the mean, median (Eq. \ref{L3_median}, Table \ref{formulae_L3}) and mode (most probable value, eq. \ref{L3_mode}, Table \ref{formulae_L3}) are all different.
Also, note that $\mu$ is not directly related to any of them, it is the inflexion point of the exponent.
Calculating the mean of the $L_3$ IMF involves incomplete Beta functions\footnote{
The incomplete Beta function is available in many scripting languages for data processing ({\sc r} (open source), {\sc idl} etc.) or via Numerical recipes \citep{PressTeukolskyVetterling-2007a}.
Sometimes what is called ``incomplete Beta function'' is actually the regularised incomplete Beta function, $I_x (p,q) = B(x;p,q)/B(p,q)$.
This is the case for the functions {\sc pbeta} in {\sc r} and {\sc ibeta} in {\sc idl}.
}
 ($B(x;p,q) = \int_0^x t^{p-1} (1-t)^{q-1} \d t$).
Using the transformation 
\< t (m) = \frac{ \( \frac{m}{\mu} \)^{1-\alpha} }{1+ \( \frac{m}{\mu} \)^{1-\alpha} } \>
the mean can be expressed as
\< E(m) &=& \mu (1- \beta)  \frac{B \(t(m_u);a,b\) - B \(t(m_l);a,b\)}{G(m_u) - G(m_l)}, \label{mean_L3} \>
where $a=\frac{2-\alpha}{1-\alpha}$ and $b = \beta - \frac{2-\alpha}{1-\alpha}$ and $G(m)$ is the auxiliary function given in Table \ref{formulae_L3}, eq. \ref{L3_aux}.

The ``peak'' of the IMF refers to the maximum in the logarithmic description.
The also very simple formula for $m_\text{P}$ is given in eq. \ref{L3_peak}, Table \ref{formulae_L3}).

\section{``Canonical'' Parameters for the $L_3$ IMF}\label{sec_std_IMF}
Observationally, the shape of the IMF is constrained mainly by the number ratios of different mass ranges to each other, for example the ratio of high-mass to low-mass stars.
Thus, a first approach to find the ``canonical'' parameters for the $L_3$ IMF could be a fit to the cumulative distributions of the Kroupa or Chabrier IMF.
This could be done in some objective way, for example by matching histograms of $L_3$ to Kroupa or Chabrier.
However, there are more properties that a ``canonically'' parametrised IMF should fulfil: 
Not only the number ratios, but also the mass ratios, the shape and the exponent should agree with each other.
We could not find an ``objective'' procedure that would fit these constraints such that for all of them the fit is good, the high-mass power-law tail leads to problems.
Therefore we choose the parameters ``by hand'' for an  optimal agreement of the $L_3$ with Kroupa and Chabrier in all the criteria.

For observational data objective fits are, of course, possible, for example with the maximum likelihood method.
There not only the upper mass exponent and the lower-mass exponent, but also the scale parameter $\mu$ can be estimated.
This is an advantage compared to the piecewise defined IMFs, where typically the ``breakpoints'' are not estimated.
It is also possible to estimate the limits, in particular $m_u$, which can also vary between star forming regions (cf. e.g. \citealp{WeidnerKroupa-2006a},
\citealp{MaschbergerClarke-2008a}, or \citealp{WeidnerKroupaBonnell-2010a} for an observational perspective and \citealp{MaschbergerClarkeBonnell-2010a} for a varying $m_u$ in simulations).

In order to normalise the IMFs to be able to find the ``canonical'' we choose $m_l = 0.01\ \Msun$, near the deuterium burning limit.
We set $m_u=150\ \Msun$, as this is commonly assumed \citep[cf.][]{WeidnerKroupa-2004a, OeyClarke-2005a,Figer-2005a}, but are aware that in some star forming regions $m_u$ can be at much higher masses \citep{CrowtherSchnurrHirschi-2010a}. 
As $m_u$ lies well in the power-law tail, the exact value of it does not affect the parameter determination.
$\alpha$, $\beta$ and $\mu$ are mainly constrained by the behaviour of the IMF below $m_\alpha$.

\subsection{$L_3$ single star IMF}
\begin{figure}
\includegraphics[width=8cm]{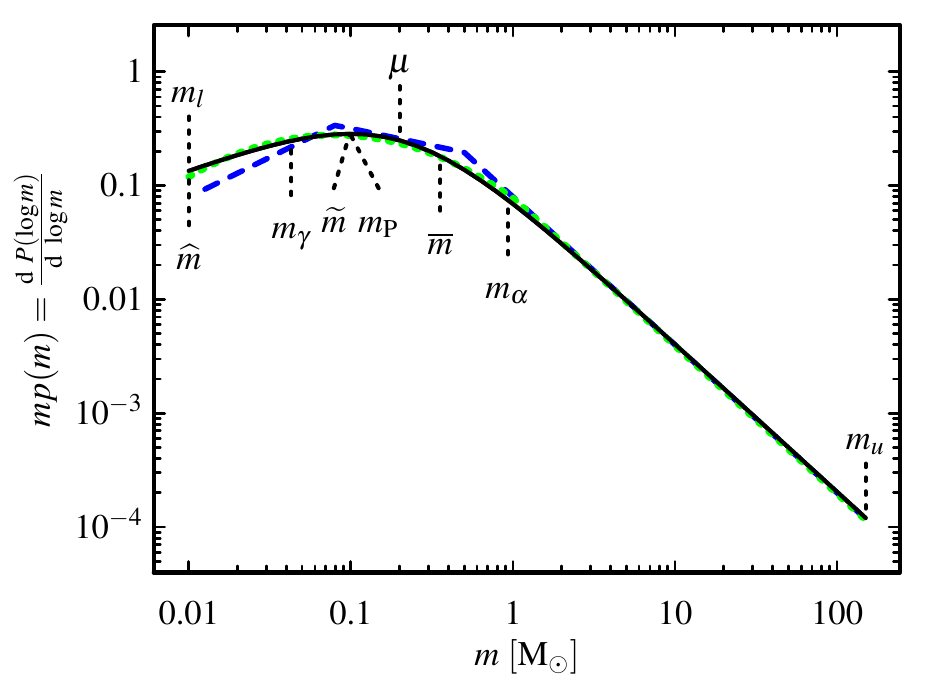}
\caption{\label{figure_imf_loglog_L3}
Probability density function for the $L_3$ IMF, like Figure \ref{figure_imf_L3} but using the logarithmic description, shown together with the \citet{Kroupa-2001a,Kroupa-2002a} IMF (dashed blue line) and the \citet{Chabrier-2003b} IMF (green dotted line).
}
\end{figure}

\begin{figure}
\begin{center}
\includegraphics[width=6cm]{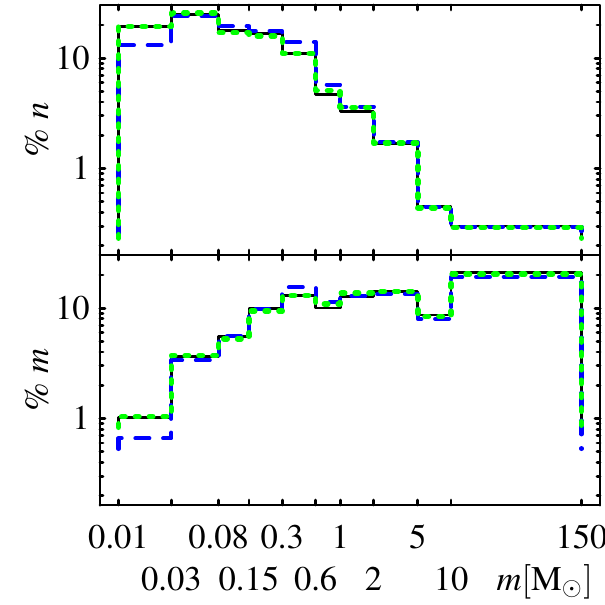}
\end{center}
\caption{ \label{figure_CDF_MDF_L3}
Comparison of the fractions of number (upper panel) and fractions of mass (lower panel) for the $L_3$ IMF (solid), \citet{Kroupa-2001a,Kroupa-2002a} IMF (blue dashed) and \citet{Chabrier-2003b} IMF (green dotted).
The $L_3$ IMF agrees better with the \citet{Chabrier-2003b} IMF, except for the mass range around 1 \Msun, where the power law is mounted onto the lognormal in the \citet{Chabrier-2003b} IMF.
}
\end{figure}

In Fig. \ref{figure_imf_loglog_L3} we show in the logarithmic description the $L_3$ IMF with parameters chosen such that it fits the ``canonical'' single-star IMF ($\alpha=2.3$, $\beta=1.4$ and $\mu = 0.2\ \Msun$).
For comparison we also show the \citet{Kroupa-2001a,Kroupa-2002a} IMF and the \citet{Chabrier-2003b} IMF, both also normalised as probabilites.
The difference between $L_3$ and  \citet{Chabrier-2003b} is marginal, between $L_3$ and \citet{Kroupa-2001a,Kroupa-2002a} equal to the difference between \citet{Kroupa-2001a,Kroupa-2002a} and \citet{Chabrier-2003b}.
The effective low-mass exponent is $\gamma = 0.48$ for $m < m_\gamma = 0.042 \ \Msun$.
The high-mass break occurs at $m_\alpha = 0.93 \ \Msun$, comparable to the start of the high mass power law of \citet{Chabrier-2003b} at 1 \Msun.
In the \citet{Kroupa-2001a,Kroupa-2002a}  IMF the high-mass power law continues to 0.5 \Msun.

A comparison of the number fraction of stars in several mass bins is shown in the top panel of Fig. \ref{figure_CDF_MDF_L3}.
The agreement between $L_3$ and \citet{Chabrier-2003b} is again very good.
The fraction of stars in the mass range 0.6--2\ \Msun\ is slightly smaller for $L_3$, because of the smooth transition to the high-mass power law.
There are differences between $L_3$ and the \citet{Kroupa-2001a,Kroupa-2002a} IMF at 0.3--1\Msun\ 
and at 0.01 -- 0.3 \Msun, caused by the segments in the Kroupa form.
The lower panel of Fig. \ref{figure_CDF_MDF_L3} shows the fraction of total mass in the mass bins,
\< \% m &=& 100 \frac{\int_{m_a}^{m_b} m p(m) \d m}{\int_{m_l}^{m_u} m p (m) \d m} \>
($m_a$ and $m_b$ being the bin limits).
$L_3$ again agrees very well with \citet{Chabrier-2003b} and well with \citet{Kroupa-2001a,Kroupa-2002a}.

As a last point we compare the exponent of the $L_3$ IMF with \citet{Kroupa-2001a,Kroupa-2002a} and \citet{Chabrier-2003b}, see Fig. \ref{figure_slope_L3}.
Interestingly, although the probability density  function, the cumulative distribution function (fraction of stars, top  panel of Fig. \ref{figure_CDF_MDF_L3}) and the mass distribution function (fraction of mass, bottom panel of Fig. \ref{figure_CDF_MDF_L3}) of the $L_3$ IMF agree more with a  \citet{Chabrier-2003b}, the exponent of the $L_3$ IMF follows more closely the  \citet{Kroupa-2001a,Kroupa-2002a} IMF.

\subsection{$L_3$ system IMF}
\begin{figure}
\includegraphics[width=8cm]{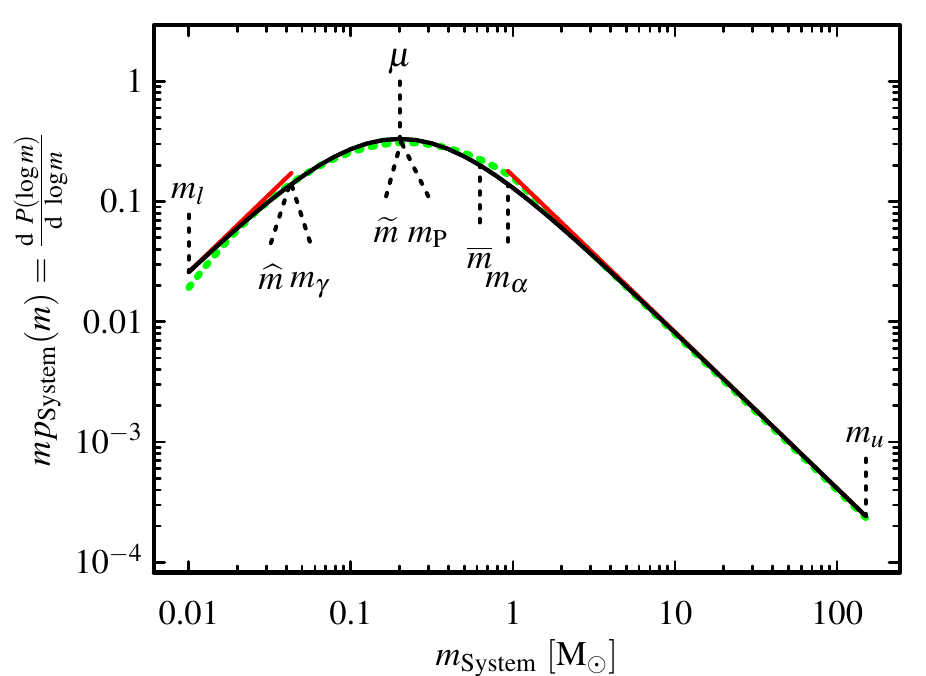}
\caption{ \label{figure_imf_system_L3}
Comparison of the $L_3$ system mass function (solid)  and the  \citet{Chabrier-2003b} system mass function (green dotted) in the logarithmic description.
}
\end{figure}

\begin{figure}
\begin{center}
\includegraphics[width=6cm]{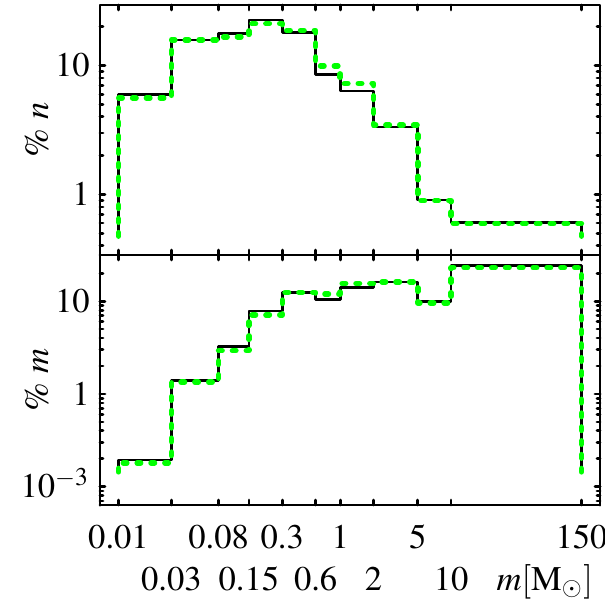}
\end{center}
\caption{ \label{figure_CDF_MDF__system_L3}
Comparison of the fractions of number (upper panel) and fractions of mass (lower panel) for the $L_3$ system IMF (solid)  and \citet{Chabrier-2003b} system IMF (green dotted).
As for the single-star IMF (Fig. \ref{figure_CDF_MDF_L3}) the lognormal-power law transition of the \citet{Chabrier-2003b} IMF around 1 \Msun\ cannot be fitted exactly.
}
\end{figure}

The system IMF for $m < 1\ \Msun$ has been given by \citet{Chabrier-2003b},
\< p_\text{Chabrier 2003, System} (m) =  
 A 0.086 \frac{1}{m} e^{-  \frac{1}{2} \(  \frac{ \log_{10} m - \log_{10} 0.22}{0.57}\)^2 },
\label{IMF_chabrier_system}
\>
and, with slightly modified parameters by \citet{Chabrier-2005a},
\< p_\text{Chabrier 2005, System} (m) =  
 A 0.076 \frac{1}{m} e^{-  \frac{1}{2} \(  \frac{ \log_{10} m - \log_{10} 0.25}{0.55}\)^2 }.
\>
($A$ is a normalisation constant).
Above $1\ \Msun$ the system IMF follows a power law with exponent $2.35$ both in \citet{Chabrier-2003b} and \citet{Chabrier-2005a}.
We adopt the \citet{Chabrier-2003b} form for $m <1\ \Msun$  and a power law with exponent $2.3$

The best parameters for $L_3$ to fit the \citet{Chabrier-2003b} {\it system} IMF are $\alpha=2.3$, $\beta=2$ and $\mu=0.2\ \Msun$, taking $m_l=0.01\ \Msun$ and $m_u = 150\ \Msun$.
A graph of both IMFs in the logarithmic description is given in Figure \ref{figure_imf_system_L3}, where very good agreement is achieved.

The effective low-mass exponent is then $\gamma = -0.3$ with breakpoint $m_\gamma = 0.043\ \Msun $ 
and high-mass breakpoint at $m_\alpha=0.93 \ \Msun.$
The mean mass is $0.62\ \Msun$ which compares well with the $0.64\ \Msun$ for the \citet{Chabrier-2003b}  system IMF.
The mass for the median (0.21\ \Msun), the ``peak'' (0.20\ \Msun) and the mass scale parameter ($\mu=0.20\ \Msun$), by chance, coincide.
Another coincidence is the near-equality of $m_\gamma$ and the mode ($\hat{m} = 0.042\ \Msun$).

As for the single star IMF, the fraction of stars and the fractions of mass over the range of mass bins is very comparable for the $L_3$ system IMF and the \citet{Chabrier-2003b} system IMF (Figure \ref{figure_CDF_MDF__system_L3}).

\section{Summary}\label{sec_summary}
The $L_3$ IMF, a functional form of the IMF generalising the log-Logistic distribution, describes the whole stellar mass range with a minimum number of parameters (3 shape, 2 limits, see Table \ref{formulae_L3} that collects all formulae).
It consists of a low-mass and a high-mass power law that are joined smoothly together.
Due to its analytical simplicity many characteristic quantities (e.g. peak and mass breaks)  can be given explicitly.
The cumulative distribution function is analytically invertible, so that drawing random masses from the $L_3$ IMF is also very simple and does not involve a large programming effort.

We have determined the parameters that fit the $L_3$ IMF to the widely used single-star IMFs of  \citet{Kroupa-2001a,Kroupa-2002a} and \citet{Chabrier-2003b} and the system IMF of \citet{Chabrier-2003b}.
The $L_3$ IMF follows these IMFs very well, obtaining the same number and mass fractions of various mass ranges, so that it is an viable alternative functional form.

\section*{Acknowledgements}
TM acknowledges funding via the ANR 2010 JCJC 0501 1 ``DESC'' (Dynamical Evolution of Stellar Clusters).
I would like to express my gratitude to Estelle Moraux for helpful discussions during writing this paper and thank Jerome Bouvier, Pavel Kroupa, J\"org Dabringhausen, Morten Anderson, Nick Moeckel and Christophe Becker for comments on the manuscript.

\bibliographystyle{mn_mod}
\bibliography{refs_tm}

\appendix
\section{The $B_4$ IMF}\label{B4_IMF}
\begin{figure}
\includegraphics[width=8cm]{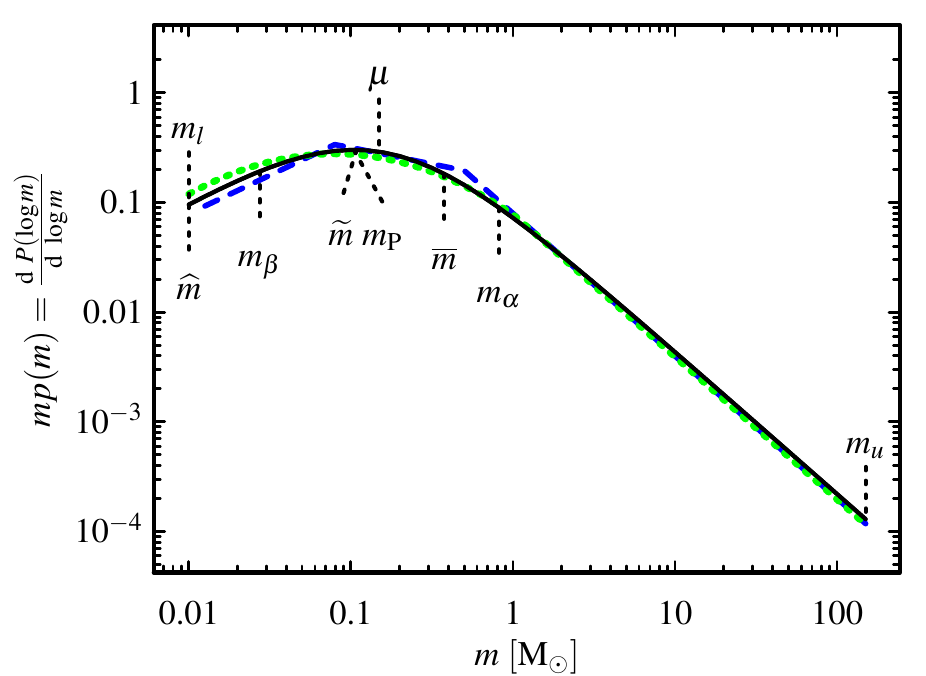}
\caption{\label{figure_imf_loglog_B4}
Logarithmic description of the $B_4$ form of the IMF, like Fig. \ref{figure_imf_loglog_L3} for $L_3$.
The blue dashed curve is the \citet{Kroupa-2001a,Kroupa-2002a} IMF and the green dotted curve is the \citet{Chabrier-2003b} IMF for comparison.
}
\end{figure}

\begin{figure}
\includegraphics[width=8cm]{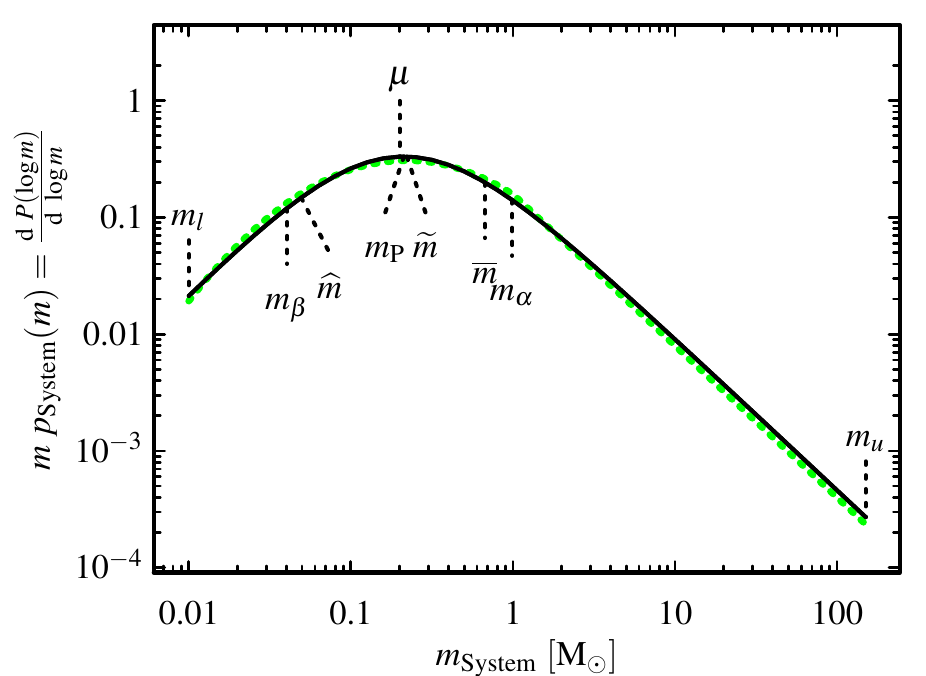}
\caption{ \label{figure_imf_system_B4}
Comparison of the $B_4$ system mass function (solid) and the \citet{Chabrier-2003b} system mass function (green dashed) in the logarithmic description, like Fig. \ref{figure_imf_system_L3}.
}
\end{figure}

In some cases it might be necessary to include explicitly the width of the IMF in its functional form. 
For this an additional parameter has to be introduced, so that the total number of parameters is 4+2, two exponents, one location and one scale parameter, plus the obligatory upper and lower mass limit.
The $L_3$ IMF can be extended to include this additional parameter, the cumulative distribution function then contains then Beta functions, thus the name $B_4$ IMF.
We give the formulae for the $B_4$ IMF in Table \ref{formulae_B4}.
The ``canonical'' parameters were determined ``manually'', as for the $L_3$ IMF, and the agreement with the IMFs for single stars \citep{Kroupa-2001a,Kroupa-2002a,Chabrier-2003b} and the IMF for systems  \citep{Chabrier-2003b} is comparably good.
Figures \ref{figure_imf_loglog_B4} and \ref{figure_imf_system_B4} show this in the logarithmic description.

\bsp

\vspace{3cm}

\begin{table*}
\caption{\label{formulae_B4}
Collection of formulae for the $B_4$ form of the IMF.
The values given for the parameter are to match the ``canonical'' single-star IMF \citep{Kroupa-2001a,Kroupa-2002a,Chabrier-2003b}, values in parentheses for the ``canonical'' system (binary star) IMF \citep{Chabrier-2003b}.
$B (t;p,q)$ is the incomplete Beta function.
}
\setcounter{BEQCounter}{0}

\begin{tabular}{clll}
\numberBEQ{B4_aux}&
Auxilliary function: &
$ \displaystyle \tilde{B}_i (m) = 
%  B_0^\infty \( \(\frac{m}{\mu}\)^{\frac{1}{\sigma}}; \sigma (\beta + i) , \sigma (\alpha - i) \) =
  B \( t(m); \sigma (\beta + i) , \sigma (\alpha - i) \) 
$
&
$ t (m) = \frac{\(\frac{m}{\mu}\)^{\frac{1}{\sigma}}}{1+\(\frac{m}{\mu}\)^{\frac{1}{\sigma}}}$
\\[3em]
& \bf Quantity & \bf Formula & \bf Definition \\
& \multicolumn{3}{l}{\bf Functional form}\\
\numberBEQ{B4_CDF}&
Cumulative distribution function (CDF) &
$ \displaystyle P_{B4} (m) = \frac{
  \tilde{B}_1 (m) - \tilde{B}_1 (m_l)
}{
  \tilde{B}_1 (m_u) - \tilde{B}_1 (m_l)
}
$ &
%$ \displaystyle P (m_l) = 0$ and $P(m_u) = 1$
\\[1em]
\numberBEQ{B4_pdf}&
Probability density function (pdf) &
$ \displaystyle p_{B4}(m)= A\frac{ 
\(\frac{x}{\mu}\)^{\beta}
}{
\( 1 + \(\frac{x}{\mu}\)^{\frac{1}{\sigma}} \)^{\sigma (\alpha+\beta)}
} $ &
%$ \displaystyle p(m) = \frac{\d }{\d m} P(m) $
\\[3em]
&   &
$ \displaystyle A = 
\frac{1}{\sigma \mu}
\frac{1}{
   \tilde{B}_1 (m_u) - \tilde{B}_1 (m_l)
} $ & 
\\[1em] 
&\multicolumn{3}{l}{\bf Parameters }\\
&High-mass exponent &
$\alpha = 2.3\ (2.3)$ &
$\alpha > 1$
\\
&Low-mass exponent &
$\beta = -0.15 $  (0.4)&
$\beta > -1 $
\\
&Location Parameter &
$\mu = 0.15$  (0.20)\ \Msun&
$\mu > 0$
\\
&Scale parameter &
$\sigma = 0.85$ (0.80) & 
$\sigma > 0$ 
\\
&Lower mass limit &
$m_l = 0.01\ \Msun$ &
$m_l > 0$
\\
&Upper mass limit & 
$m_u = 150\ \Msun $ &
$m_u > 0$
\\[1em]
&\multicolumn{3}{l}{\bf Shape characterising quantities }\\
\numberBEQ{B4_mbeta} &
Lower power-law mass limit &
$ \displaystyle m_\beta = \mu e^{-2\sigma} $ & 
$ \displaystyle p(m \in [m_l \dots m_\beta]) \approx m^{\beta}$
\\
\numberBEQ{B4_alpha} &
Upper power-law mass limit &
$ \displaystyle m_\alpha = \mu e^{2 \sigma} $ & 
$ \displaystyle p(m \in [m_\alpha \dots m_u]) \approx  m^{-\alpha}$ % \stackrel{\text{D}}{\approx}
\\
\numberBEQ{B4_slope} &
Slope (N.B. $S(\infty) = + \alpha$ {\tiny $(=  2.35)$})  & 
$ \displaystyle S(m) = - \beta + (\alpha+\beta) \frac{ \(\frac{x}{\mu}\)^{\frac{1}{\sigma}} }{ 1 + \(\frac{x}{\mu}\)^{\frac{1}{\sigma}}  } $ & 
$ \displaystyle S(m) = - \frac{\d \log p(m)}{\d \log m} $
%= - m \frac{\d \log p(m)}{\d m}  $
\\[1em]
&\multicolumn{3}{l}{\bf Scale characterising quantities }\\
\numberBEQ{B4_mean} &
Mean mass (Expectation value) &
$ \displaystyle E(m) \( = \bar{m} \)= 
\mu \frac{
   \tilde{B}_2 (m_u) - \tilde{B}_2 (m_l)
}{
   \tilde{B}_1 (m_u) - \tilde{B}_1 (m_l)
}
$ &
$ \displaystyle E(m) = \int_{m_l}^{m_u} m p (m ) \d m$
 \\[1em]
\numberBEQ{B4_median} &
Median mass &
$\tilde{m} = P^{-1} \(\frac{1}{2}\)$ (no closed form) &
$ \displaystyle P(\tilde{m}) = \frac{1}{2} $
\\[1em]
\numberBEQ{B4_mode} &
Mode (most probable mass)
&
$\hat{m} = \mu \(\frac{\beta}{\alpha} \)^{\sigma}
$ ($\beta > 0$) or $\hat{m} = m_l$ ($\beta < 0$)
&
$ \displaystyle \hat{m} = \underset{m}{\mathrm{arg\ max}}\ p(m) $
\\[1em]
\numberBEQ{B4_peak} &
``Peak'' (maximum in log-log) 
&
$ \displaystyle m_P = \mu \(\frac{\beta+1}{\alpha-1} \)^{\sigma}$ &
$ \displaystyle \frac{\d \log \( m p(m) \)}{\d \log m} = 0 $
% \left[ = m \frac{\d \log \( m p(m) \)}{\d m} \right]
\\
\end{tabular}

%\begin{tabular}{lll}
%Cumulative distribution function (CDF) &
%$ \displaystyle P(m) = \frac{
%  B \( \(\frac{x}{\mu}\)^{\frac{1}{\sigma}}; \sigma (\beta + 1) , \sigma (\alpha - 1) \)
%-  B \( \(\frac{x_l}{\mu}\)^{\frac{1}{\sigma}}; \sigma (\beta + 1) , \sigma (\alpha - 1) \)
%}{
%  B \( \(\frac{x_u}{\mu}\)^{\frac{1}{\sigma}}; \sigma (\beta + 1) , \sigma (\alpha - 1) \)
%-  B \( \(\frac{x_l}{\mu}\)^{\frac{1}{\sigma}}; \sigma (\beta + 1) , \sigma (\alpha - 1) \)
%}
%$ &
%\\
%Normalisation constant &
%$ \displaystyle A = 
%\frac{\sigma}{\mu}
%\frac{1}{
%  B \( \(\frac{x_u}{\mu}\)^{\frac{1}{\sigma}}; \sigma (\beta + 1) , \sigma (\alpha - 1) \)
%-  B \( \(\frac{x_l}{\mu}\)^{\frac{1}{\sigma}}; \sigma (\beta + 1) , \sigma (\alpha - 1) \)
%} $ & 
%\\[1em] 
%Mean mass (Expectation value) &
%$ \displaystyle E(m) \( = \bar{m} \)= 
%l \frac{
% B \( \(\frac{x_u}{\mu}\)^{\frac{1}{\sigma}}; \sigma (\beta + 2) , \sigma (\alpha - 2) \)
%-  B \( \(\frac{x_l}{\mu}\)^{\frac{1}{\sigma}}; \sigma (\beta + 2) , \sigma (\alpha - 2) \)
%}{
%  B \( \(\frac{x_u}{\mu}\)^{\frac{1}{\sigma}}; \sigma (\beta + 1) , \sigma (\alpha - 1) \)
%-  B \( \(\frac{x_l}{\mu}\)^{\frac{1}{\sigma}}; \sigma (\beta + 1) , \sigma (\alpha - 1) \)
%}
%$ &$ \displaystyle E(m) = \int_{m_l}^{m_u} m p (m ) \d m$
%\end{tabular}

\end{table*}

\label{lastpage}

\end{document}